\documentstyle[twocolumn,aps,jpc]{revtex}  
\input epsf.sty
\begin{document}

% Definition of title page:

\title{Molecular Realism in
 Default Models for Information Theories of Hydrophobic Effects
 }

\author{ M. A. Gomez\footnote{gomez@lanl.gov}, 
L.  R.  Pratt\footnote{lrp@lanl.gov}
, G. Hummer\footnote{hummer@lanl.gov}, and S.  Garde\footnote{garde@lanl.gov} 
 } 

\address{Theoretical Division, Los Alamos National Laboratory, 
Los Alamos NM 87545, USA}

\date{\today}

\maketitle

\begin{abstract}
This letter considers several physical arguments about contributions
to hydrophobic hydration of inert gases, constructs default models to
test them within information theories, and gives information theory
predictions using those default models with moment information drawn
from simulation of liquid water.  Tested physical features include:
packing or steric effects, the role of attractive forces that lower
the solvent pressure, and the roughly tetrahedral coordination of
water molecules in liquid water.  Packing effects (hard sphere default
model) and packing effects plus attractive forces (Lennard-Jones
default model) are ineffective in improving the prediction of
hydrophobic hydration free energies of inert gases over the previously
used Gibbs and flat default models.  However, a conceptually simple
cluster Poisson model that incorporates tetrahedral coordination
structure in the default model is one of the better performers for
these predictions.  These results provide a partial rationalization of
the remarkable performance of the flat default model with two moments
in previous applications.  The cluster Poisson default model thus will
be the subject of further refinement.  LA-UR-98-5431
\end{abstract}

\section{Introduction}
The idea of constructing an information theory description of cavity 
formation in water~\cite{Hummer:96} has reinvigorated the molecular 
theory of hydrophobic 
effects~\cite{Berne:96,Crooks:97,Mountain,Wallqvist,Arthur:98,Lum:98}.  
One advantage of this approach is that simple physical hypotheses can 
be expressed in a default model.  Given a fixed amount of specific 
information, the quality of the predictions gives an assessment of the 
physical ideas that are embodied in the underlying default model.  
Relevant physical ideas include: whether a direct description of dense 
fluid packings significantly improves the predictions; or whether 
incorporation of de-wetting of hydrophobic surfaces is required; or 
whether specific expression of the roughly tetrahedral coordination of 
water molecules in liquid water is the most helpful next step for 
these theories.  It is remarkable that the previous successes of the 
information models for the primitive hydrophobic effects have not 
required specific consideration of these physical points.

This letter considers these physical arguments, constructs default
models to test them, and gives the results of information theory
predictions using those default models with specific moment
information drawn from simulation of liquid water.  Occupancy moments
are used as information. Complete moment information produces results
that are independent of the default model.  However, the goal is to
judge the default models and the physical ideas that they express.
Therefore, our judgements will center on the accuracy of the
predictions with limited moment information.  More specifically, we
take the view that the quality of the prediction with two moments is
critical because information for the first two occupancy moments --
the mean and variance -- is available from experiment.

Much of the technical work required to construct the default models
considered involves molecular simulation calculations for fiducial
systems.  That technical work will be reported at a later time.

The application of the information theory approach more broadly than
to liquid water immediately turns-up cases where it works less well.
Thus, a broader suite of default models will clearly be a key
ingredient to the broader utility of this approach.

\section{Testing Physical Ideas of Hydrophobic Effects}
The information theory approach studied here grew out of earlier
studies of formation of atomic sized cavities in molecular
liquids~\cite{Pohorille:90,Pratt:91,Pratt:92,Pratt:93}. It has led to
new and simple views of entropy convergence in hydrophobic
hydration~\cite{Garde:96} and of pressure denaturation of
proteins~\cite{Hummer:98a}.  A review of these developments has been
given~\cite{Hummer:98b}; broader discussions are also
available~\cite{Pratt:98a,Pratt:98b}.

The objective of the information theory prediction is the interaction
part of the chemical potential of a hard core solute
$\beta\Delta\mu=-\ln p_0~$, where $p_0$ the probability that the hard
solute could be inserted into the system without overlap of van der
Waals volume of the solvent; 1/$\beta$=k$_B$T.  This procedure depends
on a default model $\hat{p}_{n}$ of the distribution ${p}_{n}$ 
of which
$p_0$ is the $n=0$ member.  Two default models have been considered in
previous work: (a) the `Gibbs default model' $\hat{p}_{n} 
\propto 1/n!$
that predicts a Poisson distribution when the moment $<${\it n}$>_0$ is the
only information available; and (b) the `flat default model'
$\hat{p}_{n}$ = constant$>$0, $n$ = 0, 1, $\ldots$, $n_{max}$, and zero
otherwise.  The predictions obtained using these default models for
the hydration free energy of inert gases in water are similar.
Convergence to the correct result is non-monotonic with increasing
numbers of binomial moments $B_{j} = \langle {{n} \choose {j}} 
\rangle_0$
used~\cite{Hummer:98b,Pratt:98b}. Because of this non-monotonic
convergence, the most accurate prediction obtained with a small number
of moments utilizes only the two moments, $B_1$ and $B_2$.
Furthermore, the flat default model produces a distinctly more
accurate prediction of this hydration free energy when only two
moments are used than does the Gibbs default model.

The accuracy of the prediction utilizing the flat default model is
remarkable.  Furthermore, the Gibbs default model is conceptually more
natural in this framework. So, the effectiveness of the flat default
model relative to the Gibbs model is additionally puzzling.  The work
that follows addresses these issues.

It deserves emphasis that the overall distribution $p_{n}$ is well
described by the information theory with the first two moments, $B_1$
and $B_2$.  It is the prediction of the extreme member $p_0$ that
makes the differences in these default models significant.

\subsection{Packing} 
A first idea is that the default model should contain a direct
description of dense fluid packings that are central to the theory of
liquids~\cite{WCA:83}. Accordingly, we computed $p_{n}$ for the fluid of
hard spheres of diameter d = 2.67~\AA\ at a density $\rho d^3$ =
0.633.  Those computations used specialized importance sampling and
will be reported later.  Typical predictions for the hydrophobic
hydration free energies of atomic size solutes obtained using those
results as a default model are shown in Fig.~\ref{fig2}.  That shows
the non-monotonic convergence with increasing number of occupancy
moments obtained from the flat and the Gibbs default models. The
predictions obtained using the hard sphere results as a default model
are different but not improved in the essential aspects.  Direct
convergence is only seen if four or more moments are included.  Though
the convergence is more nearly monotonic from the beginning, the
prediction obtained from a two moment model is worse than for the
flat and the Gibbs default cases.

\subsection{Attractive Interactions among Solvent Molecules} 
\begin{figure}[htbp] 
\hspace{1.5in} \epsfbox{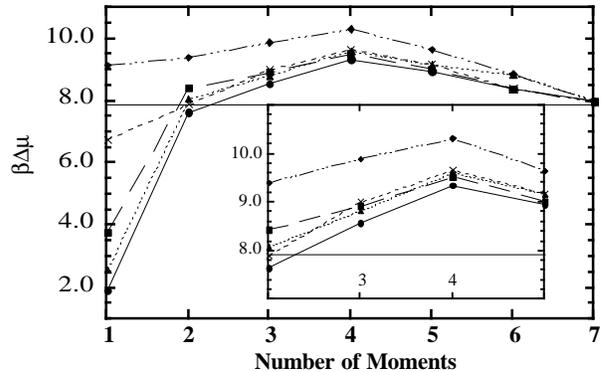} \vspace{0.1in}

\protect\caption{Convergence with number of binomial moments of 
$\beta \Delta \mu$ predicted using several default models for a
spherical solute with distance of closest approach $\lambda$ =
3.0~\AA\ for water oxygen atoms.  Identifications are: diamonds
(dash-dot lines), hard sphere default; crosses (short dash line),
Lennard-Jones default; squares (long dash line), Gibbs default;
triangles (dotted line), cluster Poisson default; circles (gray line),
flat default. For this value of $\lambda$, binomial moments $B_j$ are
non-zero through $j=9$.  The horizontal line is the prediction with all
nine moments included.  With only two moments the Lennard-Jones 
default model makes the
best prediction.
However, the differences are slight with the exception of the 
hard sphere model.
The results for the hard sphere default model were obtained from NPT
Monte Carlo calculations at $\beta$ p$^*$ = 2.989.  The average
density was $\rho d^3=0.633\pm0.002$ with d = 2.67~\AA.  The Lennard-Jones
model was obtained from NPT Monte Carlo calculations at p=0.0 and
T$^*$=1.103.  $\sigma$ = 2.67~\AA\ and the value $\epsilon /
k_{B}$=272~K was obtained from a fit of zero pressure data
\protect\cite{MacDonald}.  The mean density was $\rho^*=0.624\pm0.001$.  
A system size of 256 particles was sufficient for both simulations.  }

\label{fig2}

\end{figure}

A next idea is that attractive forces between solvent molecules might
play a significant role for these properties because attractive forces
lower the pressure of the solvent.  Dehydration of hydrophobic
surfaces becomes a principal consideration for solutes larger in size
than the solvent molecules.  But perhaps such effects are being felt
already for atomic solutes.  Accordingly, we computed $p_{n}$ for the
Lennard-Jones liquid studied by Pratt and Pohorille~\cite{Pratt:92}
for which attractive interactions were adjusted so that the
macroscopic pressure of the solvent would be approximately zero.  This
Lennard-Jones system thus gives a cavity formation free energy for
atomic sized cavities that is about the same as that of common liquid water
simulation models. The results of Fig.~\ref{fig2} confirm this latter
point but also show that the convergence with number of moments is
again non-monotonic and not better than for the flat and the Gibbs
default models. Again, direct, non-monotonic convergence is only seen
after four occupancy moments are included.

\subsection{Tetrahedral Coordination of Solvent Molecules} 
The final idea checked here is whether the predictions of cavity
formation free energies are improved by incorporating a tetrahedral
coordination structure for water molecules in liquid water.  We use a
cluster Poisson model to accomplish this\cite{NS}.  The physical
picture is: tetrahedral clusters of water molecules with prescribed
intra-cluster correlations but random positions and orientations.

\begin{figure}[tbhp]  
\hspace{1.0in}\hspace{1in}
\epsfbox{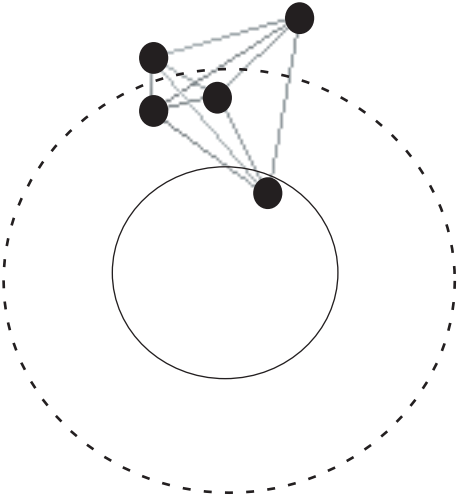} \vspace{0.1in} 

\caption{A tetrahedral cluster, the observation volume (sphere with
solid line, and the augmented volume (sphere with dashed line).  The
cluster may contribute occupants of the observation volume only if the
center is within the augmented volume.  }

\label{tetra-model}
\end{figure}

A molecular cluster may contribute to occupants of a specific
observation volume only if the center of the cluster is an occupant of
a larger augmented volume; see Fig.~\ref{tetra-model}.  Definition of
this augmented volume will depend on the structures of the clusters
and the choice of cluster center.  We then consider the generating
function $\wp(z)$ for the probability $\wp_N$ that $N$ cluster centers
are present in the augmented volume:
\begin{equation}
\wp(z) = \sum_{N=0} z^N \wp_N . \label{pi}
\end{equation}
We assume that $N$ is Poisson distributed $\wp(z) = e^{-<N>(1-z)}$
with $<${\it N}$>$ the product of the density of clusters and the volume of
the augmented region.

Next we consider the generating function $g(z)$ defined by the
conditional probabilities, $g_n$, that a cluster with center in the
augmented volume contributes $n$ oxygen atom occupants to the
observation volume:
\begin{equation}
g(z) = \sum_{n=0} z^n g_n . \label{g}
\end{equation}
Defining the generating function for the probabilities of numbers of
oxygen in the observation volume
\begin{equation}
p(z)\equiv\sum_{n=0} z^n p_n , \label{pdef}
\end{equation}
we can express
\begin{equation}
p(z) = \wp(g(z)) . \label{p} 
\end{equation}
This is a standard result of probability theory~\cite{Karlin}. $\ln
p(z)$ is a polynomial function of $z$.  Extraction of the series
coefficients from Eq.~\ref{pdef} provides the desired default model.
The numerical effort resides only in the computation of the $g_n$.

In this study, the clusters are assumed to be tetrahedra with the
oxygen atom of a water molecule at the center and at each vertex.
Thus we take $<${\it N}$>$ = $\rho$v/5, with v the volume of the augmented
region and $\rho$ the molecular density of the solvent water.  The OO
intra-cluster near-neighbor distance, the distance of a point of a
tetrahedron from its center, is 2.67\AA\ and the augmented volume is a
sphere with radius $\lambda$ + 2.67\AA.  The coefficients of $g(z)$
are obtained from a Monte Carlo calculation that randomly positions a
tetrahedron with center in the augmented volume and counts how many
O-points of the cluster land in the observation volume.

Fig.~\ref{fig2} shows the predictions for cavity formation free energy
obtained with the cluster (tetrahedron) Poisson default model. The
non-monotonic convergence is still evident. The prediction utilizing
two moments is more accurate than that utilizing the Gibbs default
model and similar to the predictions made by the flat default or the
Lennard-Jones default in the best cases considered here for those
models.

\section{Discussion} 
Each of the default models newly considered here makes specific
assumptions about n-body correlations. If the default model were the
same as the experimental distribution, the limitation of the data to
two moments would not be significant.  The optimization 
would be unaffected by the number of experimental moments
used.

The present results suggests that the efficiency of the flat and Gibbs
default models relative to the more sophisticated hard sphere and
Lennard-Jones default models might be associated with the avoidance of
specific assumptions for n-body correlations for the former cases.  In
this view, the specific assumptions for n-body correlations with the hard
sphere and Lennard-Jones default models have to
be displaced for a good description of cavity formation in liquid
water.  The third and fourth order factorial cumulants predicted on
the basis of each of these default models using two experimental
moments were evaluated and directly compared.  In fact, the
information theory predictions obtained for these moments were {\it
very\/} similar to each other.

A second point of discussion is that the biggest difference between
the Lennard-Jones and the cluster Poisson model is in simplicity.
Though the differences in the predictions seen here are not dramatic,
the cluster Poisson model is simpler.  This is particularly true for 
the dependence on thermodynamic state and the potential for further
development.  That the cluster Poisson model expressing
tetrahedral coordination appears to be a helpful new direction is
intuitive and encouraging.  However, the fact that the predictions are
not dramatically improved suggests that this sort of tetrahedral
coordination is not the only or principal physical feature relevent
for improved predictions cavity formation.

\begin{figure}[tbhp]  
\hspace{1.5in}
\epsfbox{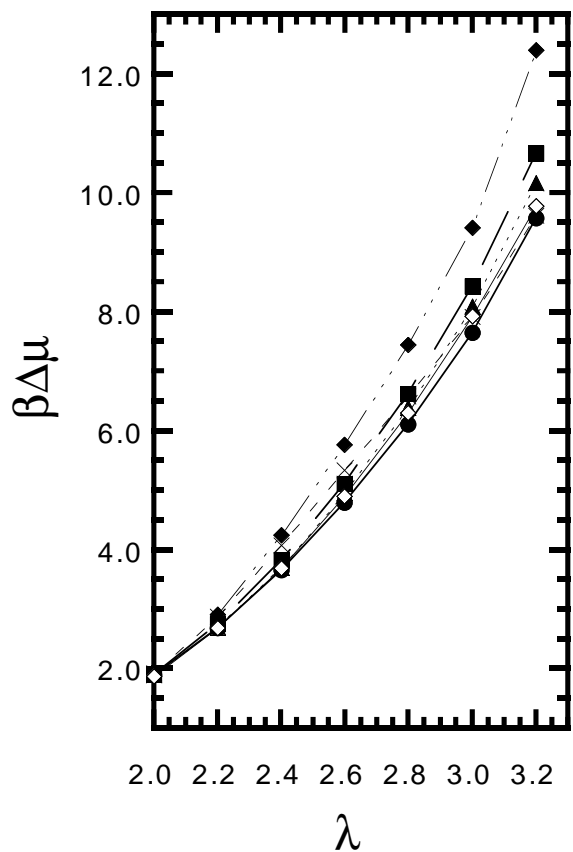} \vspace{0.1in} 

\caption{The variation of hydration free energy with solute size 
obtained with the different default models. The different models are
identified as in Fig.~\protect\ref{fig2}.  In addition, 
simulation results are shown as open diamonds.
At the smallest solute size
shown, all the models give the same result.  In the solute size range
of 2.2-2.8 \AA, the cluster Poisson model gives the best results
overall.  For larger solute sizes, the cluster Poisson model results
overestimate the hydration free energy.  At this point, results from
the Lennard-Jones default model cross the simulation results and
become slightly too small for the larger solute sizes shown.  }

\label{fig5}
\end{figure}

The Lennard-Jones default model incorporates some of the dewetting
phenomena that is expected to become more pronounced as the solute
size increases~\cite{Hummer:PRL:98}.  Fig.~\ref{fig5} shows the
variation of hydration free energy with solute size obtained with the
different default models and two moments. At the smallest solute size
shown, all the models give the same result.  In the solute size range
of 2.2-2.8~\AA, the cluster Poisson model gives the best results
overall.  For larger solute sizes, the cluster Poisson model results
overestimate the hydration free energy. At this point, results from
the Lennard-Jones default model cross the simulation results and
become slightly too small for the larger solute sizes shown.

\section{Conclusion}
We conclude that direct incorporation of dense fluid packing effects
(hard sphere default model) and packing effects plus attractive forces
that lower the pressure of the solvent (Lennard-Jones default model)
are ineffective in improving the prediction of hydrophobic hydration
free energies of inert gases over the previously used Gibbs and flat
default models.  However, a cluster Poisson model that incorporates
tetrahedral coordination structure in the default model is intuitive,
simple to implement, and is one of the better performers for these
predictions.  These results provide a partial rationalization of the
remarkable performance of the flat default model with two moments in
previous applications.  The specific cluster Poisson default model
used here is primitive and will be the subject of further refinement.

\section{Acknowledgements}  
The work was supported by the US Department of Energy under contract
W-7405-ENG-36.  The support of M.A.G by the Center for Nonlinear
Studies of Los Alamos National Laboratory is appreciated.
LA-UR-98-5431.

%\bibliography{let}

\begin{thebibliography}{10}

\bibitem{Hummer:96}
Hummer, G; Garde, S; Garcia, A.~E; Pohorille, A; Pratt, L.~R. {\em Proc. Natl.
  Acad. Sci. USA} {\bf 1996}, 93, 8951.

\bibitem{Berne:96}
Berne, B.~J. {\em Proc. Nat. Acad. Sci. USA} {\bf 1996}, 93, 8880.

\bibitem{Crooks:97}
Crooks, G.~E; Chandler, D. {\em Phys. Rev. E} {\bf 1997}, 56, 4217.

\bibitem{Mountain}
Mountain, R. D.; Thirumalai, D. {\em Proc. Nat. Acad. Sci. USA}
{\bf 1998}, 95, 8436-8440.

\bibitem{Wallqvist} Wallqvist, A.; Covell, D. G.; Thirumalai, D. 
{\em J. Amer. Chem. Soc.} {\bf 1998}, 120, 427-428.

\bibitem{Arthur:98}
Arthur, J.~W; Haymet, A. D.~J. {\em J. Chem. Phys.} {\bf 1998}, 109, 7991.

\bibitem{Lum:98}
Lum, K; Chandler, D; Weeks, J.~D (unpublished).

\bibitem{Pohorille:90}
Pohorille, A; Pratt, L.~R. {\em J. Amer. Chem. Soc.} {\bf 1990}, 112, 5066.

\bibitem{Pratt:91}
Pratt, L.~R. {\em CLS Division 1991 Annual Review} National Technical
  Information Service U. S. Department of Commerce: 5285 Port Royal Rd.,
  Springfield, VA 22161, 1991, LA-UR-91-1783.

\bibitem{Pratt:92}
Pratt, L.~R; Pohorille, A. {\em Proc. Natl. Acad. Sci. USA} {\bf 1992}, 89,
  2995.

\bibitem{Pratt:93}
Pratt, L.~R; Pohorille, A.  in {\em Proceedings of the EBSA 1992 International
  Workshop on Water-Biomolecule Interactions}, edited by Palma, M.~U;
  Palma-Vittorelli, M.~B; Parak, F. Societa  Italiana di Fisica: Bologna, 1993.

\bibitem{Garde:96}
Garde, S; Hummer, G; Garcia, A.~E; Paulaitis, M.~E; Pratt, L.~R. {\em Phys.
  Rev. Letts.} {\bf 1996}, 77, 4966.

\bibitem{Hummer:98a}
Hummer, G; Garde, S; Garcia, A.~E; Paulaitis, M.~E; Pratt, L.~R. {\em Proc.
  Natl. Acad. Sci. USA} {\bf 1998}, 95, 1552.

\bibitem{Hummer:98b}
Hummer, G; Garde, S; Garc\'{\i}a, A.~E; Paulaitis, M.~E; Pratt, L.~R. {\em J.
  Phys. Chem. B} {\bf 1998}, 102, 10469.

\bibitem{Pratt:98a}
Pratt, L.~R.  in {\em Encyclopedia of Computational Chemistry}, edited by
  Schleyer, P. v.~R; Allinger, N.~L; Clark, T; Gasteiger, J; Kollman, P.~A;
  Schaefer~III, H.~F; Schriener, P.~R. John Wiley \& Sons: Chichester, 1998.

\bibitem{Pratt:98b}
Pratt, L.~R; Garde, S; Hummer, G.  in {\em New Approaches to Old and New
  Problems in Liquid State Theory - Inhomogeneities and Phase Separation in
  Simple and Complex Systems}, {\em NATO Advanced Study Institute}, edited by
  Caccamo, C. Kluwer: Dordrecht, 1998.

\bibitem{WCA:83}
Chandler, D; Weeks, J.~D; Andersen, H.~C. {\em Science} {\bf 1983}, 220, 787.

\bibitem{NS}
Neyman, J; Scott, E.~L. {\em Scientific American} {\bf September, 1956}, 187.

\bibitem{Karlin}
Karlin, S; Taylor, H.~M. {\em A First Course in Stochastic Processes} Academic
  Press Inc.: New York, 1975.

\bibitem{Hummer:PRL:98}
Hummer, G; Garde, S. {\em Phys. Rev. Letts.} {\bf 1998}, 80, 4193.

\bibitem{MacDonald}
MacDonald, I.~R; Singer, K. {\em Molec. Phys.} {\bf 1972}, 23, 29.

\end{thebibliography}
%\bibliographystyle{prsty}
%\bibliographystyle{custome}

\end{document}